\documentclass[12pt,preprint]{aastex}

\shorttitle{SNR 0453-68.5: A Composite Remnant}
\shortauthors{McEntaffer et al.}

\begin{document}

\title{SNR 0453-68.5: An Asymmetric Remnant and its Plerion in the Large Magellanic Cloud}

\author{R. L. McEntaffer, T. Brantseg, \& Morgan Presley}
\affil{Department of Physics and Astronomy, University of Iowa, Iowa City, IA 52242}
\email{randall-mcentaffer@uiowa.edu}

\begin{abstract}
We present a comprehensive study of the X-ray emission from SNR 0453-68.5 in the Large Magellanic Cloud (LMC) as seen from the \textit{Chandra} X-ray Observatory.  This object is in a class of composite remnants that exhibit a shell of emission surrounding a central plerion, more commonly known as a pulsar wind nebula (PWN).  This is one of only five remnants in the LMC with an identified PWN.  We find that the shell of emission is not ejecta dominated, but rather due to shocked ISM that has been swept up by the supernova blast wave or located in a precursor cavity wall.  This is supported by the morphology of the local molecular cloud as seen with the \textit{Spitzer} Space Telescope.  The spectral properties are consistent with a middle-aged remnant $>$17000 years old.  A probable point source within the central knot is determined to be the pulsar powering the synchrotron emission of the PWN.  Spectral fits show the nebula is well characterized by a power law with photon index $\Gamma=2.0$.  This index is constant over a spatial scale of 0.4-1.2 pc, which is inconsistent with younger PWN containing remnants such as the Crab Nebula and SNR 0540-69.3.  These fits also contain significant contributions from an ejecta dominated thermal plasma which we interpret as evidence of mixing during an evolved interaction of the PWN with the reverse shock of the SNR.  We observe no evidence that the central pulsar contains a significant velocity transverse to the line of sight and argue that despite the asymmetric surface brightness distribution the SN explosion giving birth to this remnant may have been quite symmetric.
\end{abstract}

\section{Introduction}\label{intro}
Supernova remnants (SNRs) are important astronomical objects given their prominent role in enrichment of the interstellar medium (ISM), galactic mass and energy feedback, and stellar evolution.  SNRs display a wide variety of morphologies from circularly symmetric emission to a complex of knots and shocks in a confused medium.  Classification of the precursor explosion can be further complicated by the absence of a compact central source or if the remnant is old and confused with the local ISM.  The situation can be somewhat improved in the case of a composite remnant.  This is a class of Type II SNR that exhibits a shell of emission combined with a central plerion, also referred to as a pulsar wind nebula (PWN), surrounding a pulsar (see \citet{GaenslerSlane} for a review of PWN).  The power lost as a pulsar spins down is transferred to a wind of relativistic particles.  The nebula glows with synchrotron radiation as the particles gyrate in the nebula's magnetic field.  PWNe form a termination shock as they sweep up local ejecta.  Other morphological features can include bipolar jets, toroidal fields, and structures due to proper motion of the pulsar such as bow shocks and long tails.  The \textit{Chandra X-ray Observatory} is an ideal tool for studying these objects as the high spatial resolution combined with modest spectral resolution can distinquish between these various features.  To date, $>$50 PWNe have been observed by \textit{Chandra} (see \citet{KP2008} for an overview of these observations).  Of these, only five have been identified in the Large Magellanic Cloud (LMC), one of which is contained within the composite remnant SNR 0453-68.5 \citep{Gaensler}.  The other four are contained within SNR 0540-69.3 \citep{Park,Petre}, N206 \citep{Williams}, DEM L241 \citep{Bamba}, and N157B \citep{Chen,Dennerl,WG98}.

\citet{Gaensler} have performed a study of the radio and X-ray emission of the SNR 0453-68.5 PWN.  They report a power law fit to the high energy central core emission with a spectral index of $\alpha=-0.9\pm0.4$.  They find only marginal evidence for a point source.  Analysis of the radio nebula shows that the morphology is elongated, which the authors interpret as being indicative of an interaction between the PWN and the reverse shock of the SNR.  Furthermore, the centroids for the X-ray and radio emission overlap indicating a lack of proper motion of the pulsar.  Another broadband study performed by \citet{Williams2006} compares the \textit{Chandra} X-ray and \textit{Spitzer} IR morphologies.  The coincidence with IR and X-ray emission lead to a conclusion that the blast wave has swept up the local interstellar medium (ISM) leading to thermal dust emission.  This suggests that the X-ray emission could be dominated by these interactions with the environment and not dominated by hot ejecta.  However, no analysis has been published on the physical properties of the X-ray emitting shell which would verify this possibility.

The X-ray surface brightness distribution of SNR 0453-68.5 has been studied somewhat by \citet{Lopez2011}.  This is an analysis of the symmetry of 24 remnants in the Milky Way and LMC.  They perform their investigation using the multipole moments present in the X-ray surface brightness of the \ion{Si}{13} line (1.75--2.0 keV) in addition to images in the soft X-ray band (0.5--2.1 keV).  The sample population of remnants is chosen using young age as a discriminator which supports an assumption that these are ejecta dominated sources.  If this is the case then any asymmetries in morphology may lend insight into the precursor explosion.  Their key findings are that CC SNRs tend to be more elongated, more elliptical, and exhibit more mirror asymmetry when compared with Type Ia SNRs.  Their interpretation is that this dichotomy is driven by the different explosion mechanisms and different local environments exhibited by these classes of SNR.  \citet{Lopez2011} classify SNR 0453-68.5 as a Type II SNR based upon the detection of a central point source and apparent asymmetric morphology.  However, it is unclear whether this asymmetry is due to the precursor explosion or anisotropies in the local environment.

The composite nature and asymmetry of SNR 0453-68.5 is intriguing.  Is the asymmetry linked to the explosion mechanism?  Is this asymmetry consistent between the shell of emission and the PWN?  Has the link to the explosion mechanism been lost and the asymmetry completely due to the environment?  Is the PWN emission consistent with other LMC PWN or unique to this remnant?  A study of the morphology and physical characteristics of SNR 0453-68.5 will not only increase our understanding of this class of remnants in the LMC, but asymmetric, composite SNR in general.  The study presented here characterizes the physical properties of the shell emission from SNR 0453-68.5 and utilizes the high spatial resolution of \textit{Chandra} to clarify the structure present in this emission.  This combination of spatial and spectral data make it possible to model the interaction between the SNR blast wave and the local environment.  The morphology of the local ISM is determined through IR observations from \textit{Spitzer}.  We also present a spatial and spectral analysis of the PWN region showing that the emission is due to an evolved interaction of the PWN with the reverse shock of the SNR.  This forms the most comprehensive analysis of the X-ray emission from SNR 0543-68.5 to date, and characterizes one of only five PWN containing remnants in the LMC.

\section{Observations and Data Reduction}
SNR 0453-68.5 was observed with the Advanced CCD Imaging Spectrometer (ACIS) on the \textit{Chandra} X-ray Observatory for 40 ks. The observation was made on December 18, 2001 and archived as observation ID 1990. The entire supernova remnant was imaged on the back-illuminated S3 chip. The back-illuminated ACIS chips are more sensitive to soft X-ray emission, which makes them useful for objects such as SNR 0453-68.5 which are dominated by soft X-ray emission. The observation produced imaging with an unbinned spatial resolution of $\lesssim$1''.

The level 1 data were reprocessed to level 2 with standard processing procedures in the Chandra Interactive Analysis of Observations (CIAO, v 4.4; \citet{CIAO2006}) software package with current calibration data from the Chandra Calibration Database (CALDB, v.4.4.2). Good time intervals, charge transfer inefficiency, and time-dependent gain variations were accounted for. The extracted spectra were background-corrected using adjacent regions of the chip that were devoid of emission. 

The infrared images are taken from the \textit{Spitzer} Space Telescope.  SNR 0453-68.5 was observed with the Multiband Imaging Photometer for \textit{Spitzer} (MIPS; \citet{RiekeYoung2004}) instrument on March 2, 2005 as part of \textit{Spitzer} program ID 3680.  The observation resulted in a $\sim$31 s integration time in the 24 $\mu$m bandpass. The data were retrieved from the \textit{Spitzer} Heritage Archive website after undergoing standard pipeline processing procedures and are presented here without further processing.

\begin{figure} [htbp]
   \centering
   \includegraphics[width=6.0in,height=5.76in]{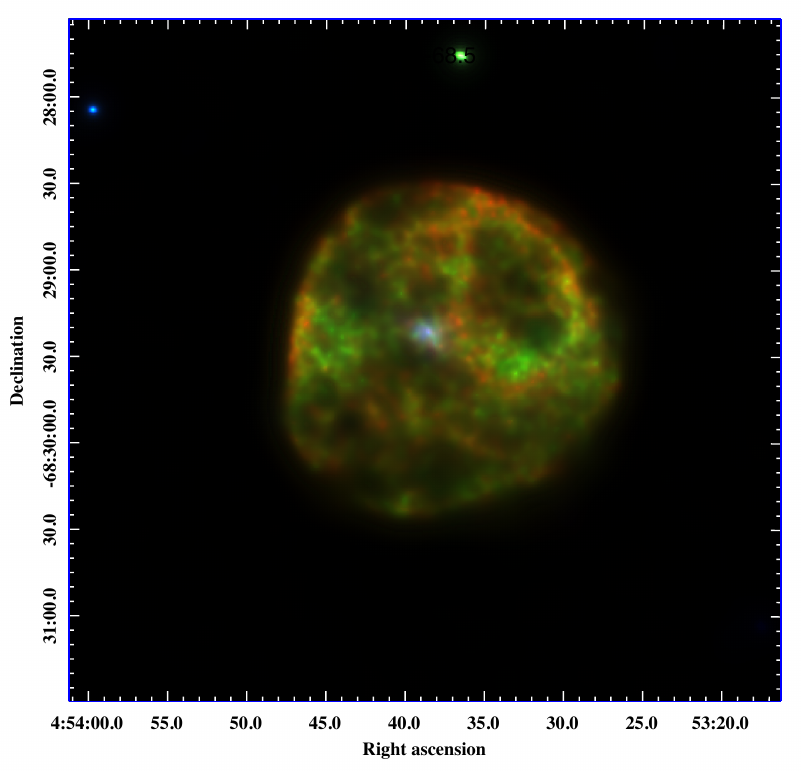}
   \caption{Image of the raw data from the \textit{Chandra} ACIS S3 back-illuminated CCD.  The data have been smoothed using the CIAO tool dmimgadapt.  This is Gaussian smoothing using a variable kernel radius from 0.1-10 pixels with an integrated count threshold of 30 per smoothing kernel.  The intensity in each color is scaled from $0\sim2.5$ counts per bin.}
   \label{raw_xray}
\end{figure}

\begin{figure} [htbp]
   \centering
   \includegraphics[width=6.0in,height=5.76in]{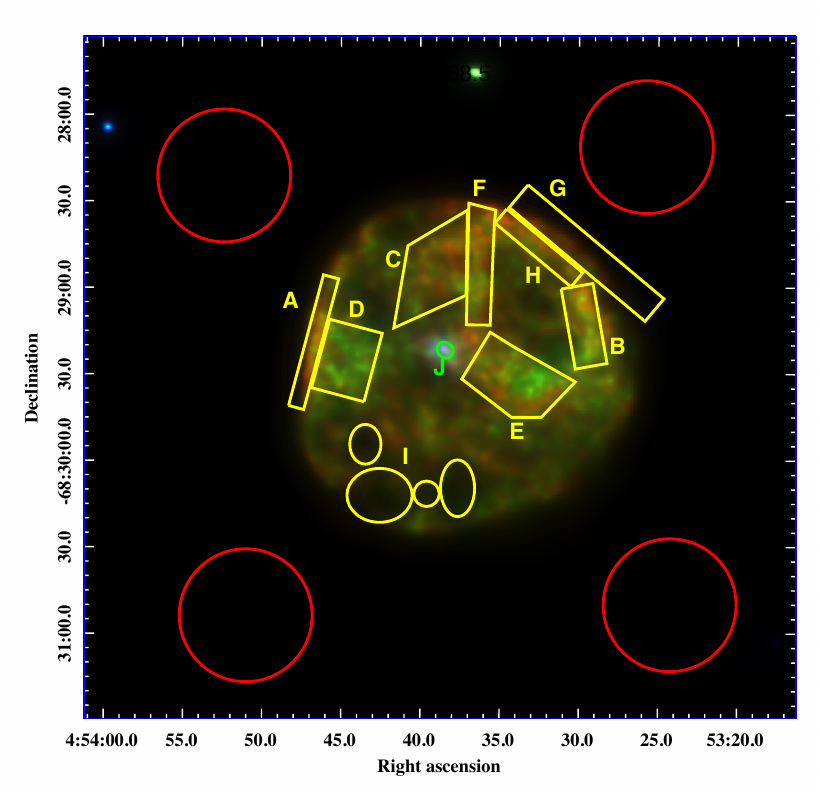}
   \caption{Spectral extraction regions A-J overlaid on the tri-color image.  The background regions are the four circles exterior to the remnant.}
   \label{regions}
\end{figure}

The reprocessed raw X-ray data are shown as a false tri-color image in Figure \ref{raw_xray}. The energy ranges corresponding to each color are 0.3--0.7 keV for red, 0.7--1.2 keV for green, and 1.2--8.0 keV for blue.  The image is created using a gaussian smoothing tool in CIAO, \textit{dmimgadapt}, with a variable radius smoothing function varying logarithmically between 0.1 and 10 pixels with a minimum count threshold of 30.  The intensity is scaled from $0\sim2.5$ counts/bin per color.  The strongest intensity lies near the center at the site of the PWN \citep{Gaensler}, which is where the majority of the high energy, blue emission is also located.  External to this there are larger areas of dimmer, more diffuse emission that is filamentary in nature.  There are also large voids of emission in the northeast and northwest quadrants as well as the southern hemisphere.  The emission in the south is much dimmer overall than the emission in the north.  The emission dies into the background very rapidly at the edge of the remnant.  This is sometimes marked by additional brightness enhancements indicative of shocked material as seen in the east, the west, and the northwest limbs.  These shock heated regions are particularly bright in soft X-ray emission.  We use these general morphological features as a basis for drawing regions used for spectral extraction and analysis.  These regions are seen in Figure \ref{regions} and are named with the letters A-J: regions A, H, and G trace shocked plasma near the limb; regions B, C, D, E, and F encompass the internal, more diffuse emission, yet are still bright; region I probes the voids in the southern hemisphere.  Finally, region J encompasses the point source, a presumed pulsar, at the center of the remnant.  This region is determined using \textit{wavdetect} in CIAO \citep{freeman}. This tool finds sources in the data set by correlating the image with ``Mexican Hat'' wavelet functions. The tool then draws an elliptical region around the detected source out to a specified number of standard deviations. In the present case, the detection scales used are 2 and 4 pixels, and the region size is 3$\sigma$.  The results from the spectral extractions of these regions are discussed in the following section.

\section{Spectral Extraction}
Data are extracted from the regions detailed in Figure \ref{regions} in order to perform spectral analyses. These regions are chosen to isolate physically distinct features, based on an inspection of the morphology in the raw data. Spectra for each of these regions are extracted using the \textit{specextract} tool in CIAO. This tool automatically creates ancillary response files and redistribution matrix files for each region.  A background is created from the four circular regions shown on the exterior of the remnant and is subtracted from each of the extracted SNR spectra.  The spectral resolution obtained by \textit{Chandra} is $\sim$5--20 ($E/ \Delta E$) over the energy band used, 0.3-1.5 keV.  Data below 0.3 keV are not used, owing to the uncertainty of the low-energy ACIS calibration, and emission above 1.5 keV is negligible in the regions selected.  The data are binned to at least 20 counts per bin to allow the use of Gaussian statistics.  To analyze the extracted spectra, we use the CIAO modeling and fitting package, Sherpa \citep{Sherpa}.

If the X-ray emission is caused by shock heating of the ejecta or ISM, we expect to see prominent emission lines. Depending on the plasma density and the time since it has been shocked, we expect the emitting material to be in either collisional ionization equilibrium (CIE; if the plasma has high density or was shocked a long time ago) or in nonequilibrium ionization (NEI).  We use the \textit{xsvnei} model for nonequilibrium conditions (\citet{bork94,bork01,Hamilton,liedahl95}.  We replace the default line list with an augmented list developed by Kazik Borkowski that includes more inner shell processes especially for the Fe-L lines.  For equilibrium conditions we use the \textit{xsvapec} model, which uses an updated version of the ATOMDB code (v2.0.1; \citet{aped1,aped2}) to model the emission spectrum.  We include a second temperature component to these fits if it is determined statistically relevant as determined from an F-test (probabilites $<0.05$ indicate a statistical improvement given the additional component).  In addition, we investigate the significance of a contribution from a non-thermal component by including the \textit{xssrcut} and$/$or \textit{xspowerlaw} models to each region \citep{RK1999,Reynolds1998}.  To account for interstellar absorption along the line of sight, these models are convolved with a photoelectric absorption model, \textit{xsphabs}. Dielectronic recombination rates are taken from \citet{mazz} with abundances from \citet{angr} and cross-sections from \citet{bcmc}.  

\begin{figure} [htbp]
   \centering
   \includegraphics[width=6.0in,height=4.54in]{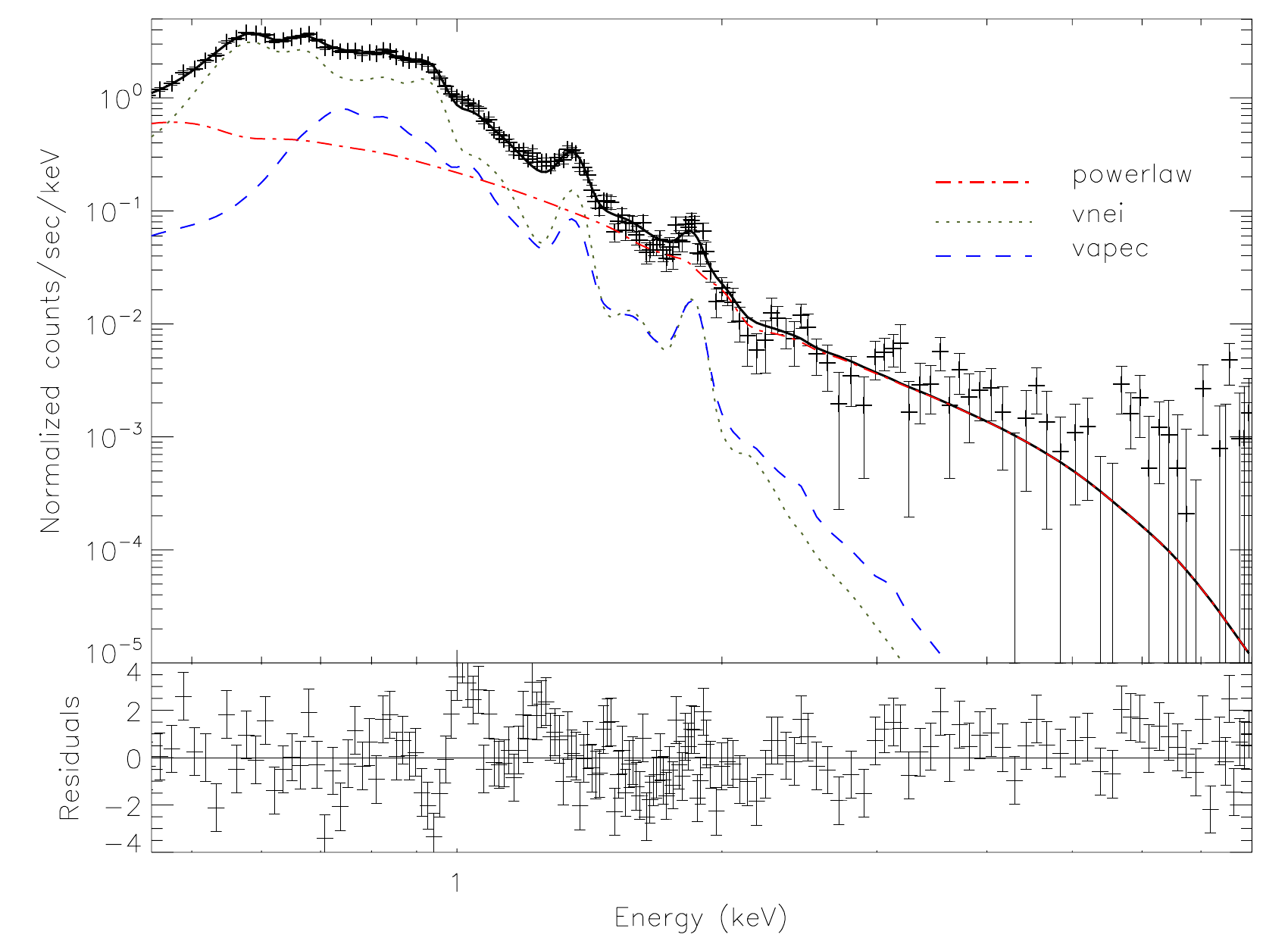}
   \caption{Spectrum extracted from the entire remnant.  The best fit model is the solid line.  Individual components include a powerlaw (dash-dotted line), a nonequilibrium plasma (dotted line), and an equilibrium plasma (dashed line).}
   \label{TotalSpectrum}
\end{figure}

The spectrum of the entire remnant is shown in Figure \ref{TotalSpectrum}.  Several attempts were made to fit this global spectrum.  The best fit consists of three components, a low temperature equilibrium plasma ($kt\sim0.13$ keV), a higher temperature nonequilibrium plasma ($kt\sim0.39$ keV), and a power law ($\Gamma\sim3.5$), with a statistic of $\chi^2/dof=217/153$, where $dof$ is degrees of freedom.  While the spectrum is confused by the many distinct regions and physical properties that are evident from the data, these components form a basis for the individual region fits where we might expect similar plasma conditions.  Ideally, each region will be small enough to isolate a physically distinct plasma while maintaining a reasonable number of counts for adequate statistics.

The variable fit parameters include the absorbing column density, $N_H$, the temperature of each component, $kT$, a normalization parameter, $norm$, the ionization timescale for any NEI components, $\tau$, and elemental abundances.  The initial values for all elemental abundances are set to those typical of the LMC as described in \citet{RD1992}. These abundances, relative to solar are: He 0.89, C 0.26, N 0.16, O 0.32, Ne 0.42, Mg 0.74, Si 1.7, S 0.27, Ar 0.49, Ca 0.33, Fe 0.50, and Ni 0.62.  Emission line lists in the 0.3--1.5 keV energy range for plasmas with temperatures $kT\sim$0.09--2.0 keV show that the emission is dominated by highly ionized states of C, N, O, Ne, and Fe with contributions from Mg and Si.  The spectral fits begin with all abundances frozen to LMC levels.  A given element is allowed to vary if it significantly improves the fit.

\section{Results}
\begin{figure} [htbp]
   \centering
   \includegraphics[width=6.0in,height=4.6in]{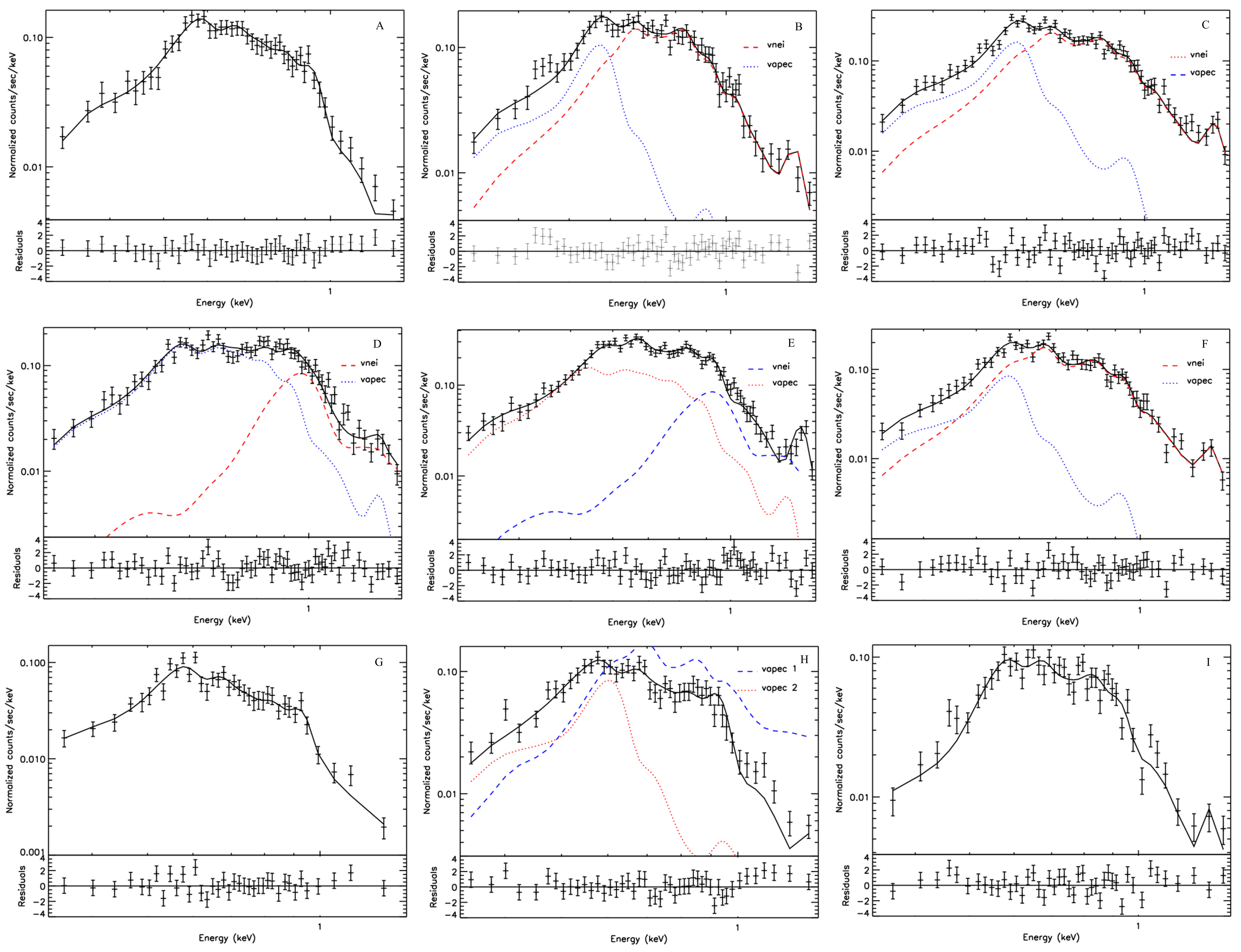}
   \caption{Spectra from the extracted regions with the best fit models shown as solid lines.  Where applicable, individual components are shown.  Residuals are shown beneath each plot.}
   \label{spectra}
\end{figure}

The best fit parameters for each region are given in Table \ref{fitparams}.  Quoted errors are 90\% confidence intervals for the parameter of interest and are calculated using the \textit{conf} tool in Sherpa.  This tool allows all thawed fit parameters to vary when calculating the change in fit statistic.  The absorbing column density is allowed to vary, yet its value stays quite close to the average LMC value observed by the LAB survey of \ion{H}{1}, $0.19\times10^{22}$ cm$^{-2}$ \citep{LAB2005}.  The ionization timescale can be represented by the product of the time since shock heating and the electron density of the shocked material ($\tau = n_{e}t$), with $\tau\gtrsim$ a few $10^{12}$ cm$^{-3}$ s usually signifying equilibrium conditions \citep{Smith2010}. This parameter is only used in the NEI models. The presence of a non-thermal plasma is tested with the addition of a synchrotron component, \textit{xssrcut} and$/$or \textit{xspowerlaw}, but these attempts did not yield better fit statistics for regions outside J than those with the thermal models alone.  The extracted spectra from each region are shown in Figure \ref{spectra}. For each region, the best-fit model is overlaid as a solid line.

\begin{figure} [htbp]
   \centering
   \includegraphics[width=6.0in,height=4.5in]{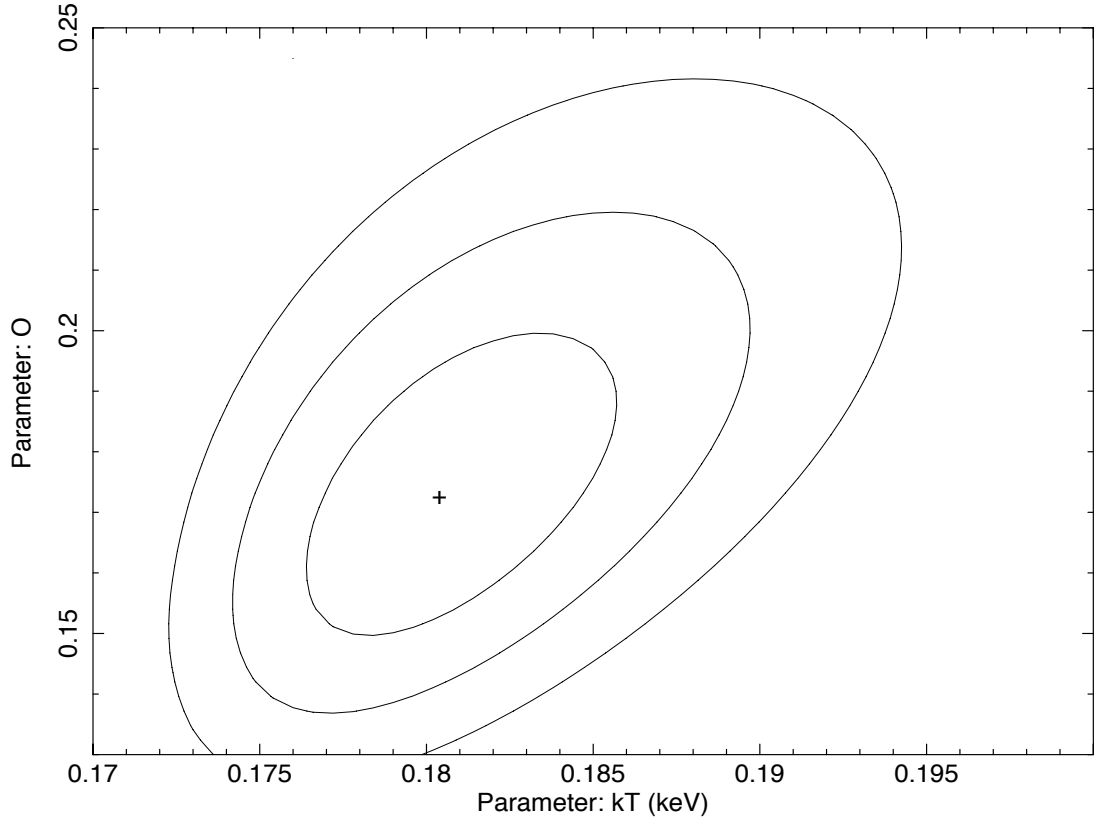}
   \caption{The contours correspond to the $1\sigma$, $2\sigma$, and $3\sigma$ changes in the $\chi^2$ statistic as the $kT$ and O abundance parameters are varied for region A.}
   \label{contours}
\end{figure}

Region A outlines a bright, shocked filament in the eastern limb of the remnant and is best fit with a single-temperature plasma in collisional ionization equilibrium.  The temperature is quite soft, $kT=0.18$ keV, and varying O is necessary to obtain the best fit.  The O abundance of 0.17 is well below that typical of the LMC, 0.32.  The 90\% confidence level errors for the temperature are very tight, thus requiring a contour map of $\chi^2$ values as a better representation of the parameter space.  Figure \ref{contours} displays contours corresponding to 1$\sigma$, 2$\sigma$, and 3$\sigma$ changes in $\chi^2$ as the parameters are varied.  This verifies the sensitivity of the region A fit to the O and temperature parameters and shows that O is underabundant and inconsistent with the LMC value of 0.32 relative to solar.  The G and H regions can also be best described by a single temperature CIE plasma.  As with region A, other models (NEI, CIE$_1+$CIE$_2$, and CIE$+$NEI) do not statistically improve the fit.  With temperatures of $kT=0.18$ keV and O abundances of 0.17 and 0.15 respectively, the plasma in these regions are nearly identical to A ($\chi^2$ contours are similar as well).  Region H displayed a slight depletion of Fe in the best fit, but not significantly below the LMC value of 0.5.  Similarly to region A, we selected regions G and H with a goal of isolating bright, linear features at the very edge of the remnant.  Regions G and H trace out two separate shocked regions that appear oriented close to one another, although projection effects may make this juxtaposition coincidental.  Even though region A is on the opposite side of the hemisphere, it exhibits plasma conditions that are quite similar to those in G and H.


Regions E and I can both be described by single temperature NEI plasmas.  Additional temperature components do not improve the quality of the fits.  In comparison to the previously discussed regions, E is located in an area of noticeably harder emission (see Figure \ref{regions}).  This dominant component drives the temperature of the fit to $kT=0.35$ keV, much higher than the other bright emission regions.  The O abundance is still depleted with a $>3\sigma$ depletion in Fe as well.  In region I, a single temperature CIE model is also able to fit the spectrum, but a F-test probability of 0.02 favors the NEI state.  Region I is unique as it traces the dimmer, hotter interior of the remnant as opposed to bright features.  This is evidenced by the much higher temperature of $kT=0.42$ keV.  The abundances in region I are consistent with those of the local ISM.  The O abundance is varied to achieve the best fit, but is not significantly below the LMC value.

The remainder of the regions, B, C, D, and F, are best fit with a two temperature component model.  One temperature models fit B, C, and D poorly, with each having a fit statistic of $\chi^2/dof\sim2$. The two temperature, CIE$+$NEI models fit these regions with $\chi^2/dof=62/51$, $\chi^2/dof=81/55$, and $\chi^2/dof=79/55$, respectively.  In region F, a single temperature model fits with $\chi^2/dof=77/51$ while a two temperature model (CIE$+$NEI) fits statistically better with $\chi^2/dof=61/49$.  The plasma conditions in these regions are very similar to each other with a higher temperature NEI component combined with a low temperature CIE plasma, both at abundances consistent with the LMC.  In each of the best fit cases the value for ionization timescale, $\tau$, is only a lower limit.  This may suggest CIE conditions and led to attempts with two temperature component CIE fits.  In regions B, C, and D the best fit double CIE models cannot constrain the emission measure for either component with $1\sigma$ errors consistent with zero emission in both regions.  Also, the two component CIE model cannot adequately fit region F ($\chi^2_{red}>2$).  Therefore, despite the lack of constraint on $\tau$, multiple component CIE plasmas do not provide a convincingly better fit.  Given the NEI conditions measured in the interior regions E and I we may expect contributions from an NEI plasma due to line-of-sight confusion in regions B, C, D, and F.  We therefore adopt the CIE$+$NEI plasma as the best fit in these regions.

\begin{figure} [htbp]
   \centering
   \includegraphics[width=6.0in,height=4.81in]{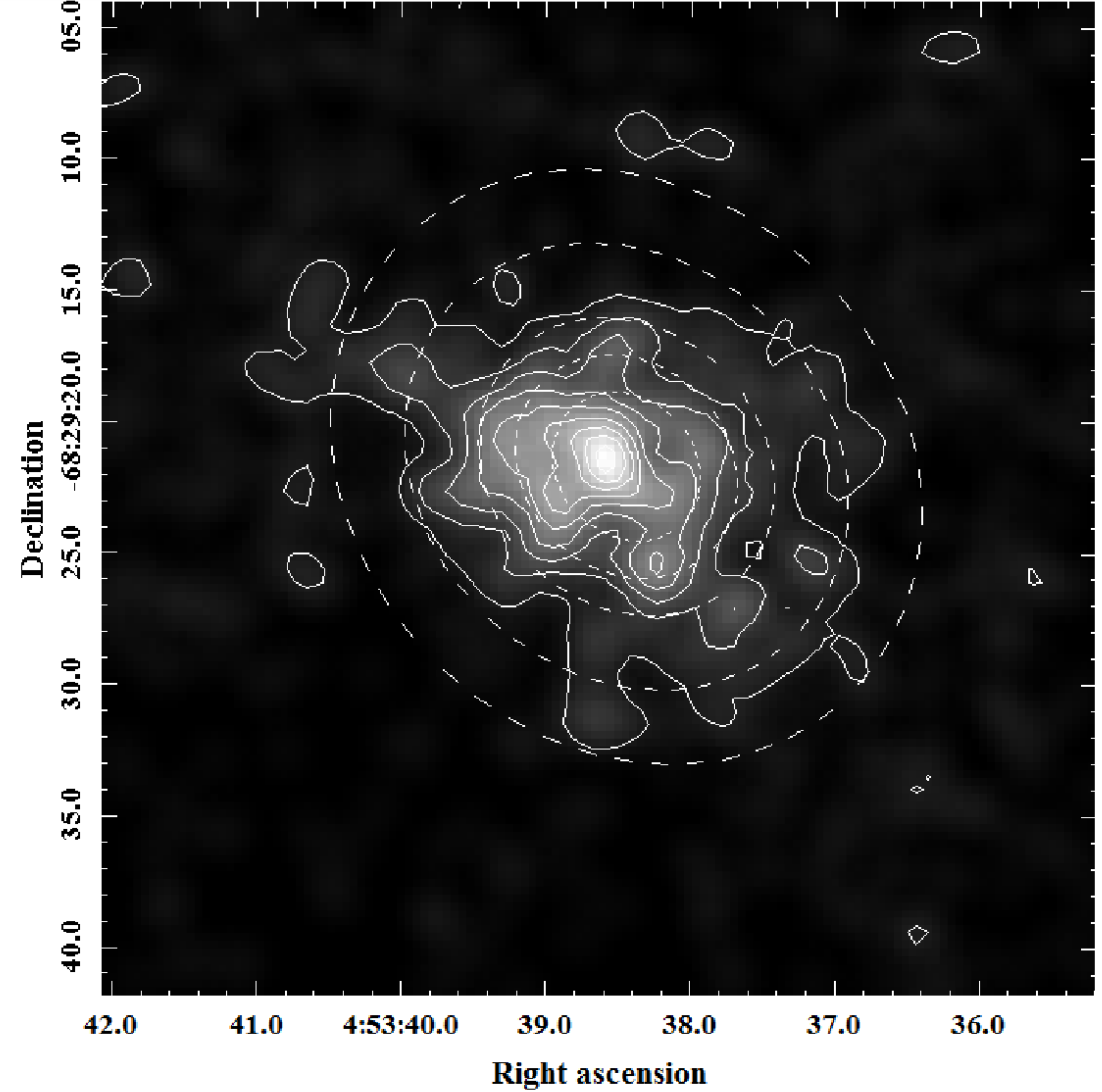}
   \caption{A zoom-in on the PWN.  Raw, unbinned data from 1.2--8.0 keV are shown in grayscale with 10 contours evenly spaced from 0--3 counts per bin.  The data have been gaussian smoothed with a 3 pixel kernel.  Dashed ellipses correspond to extraction regions J, $1.5\times$J, $2\times$J, $3\times$J and $4\times$J.}
   \label{Jpic}
\end{figure}
\begin{figure} [htbp]
   \centering
   \includegraphics[width=6.0in,height=4.5in]{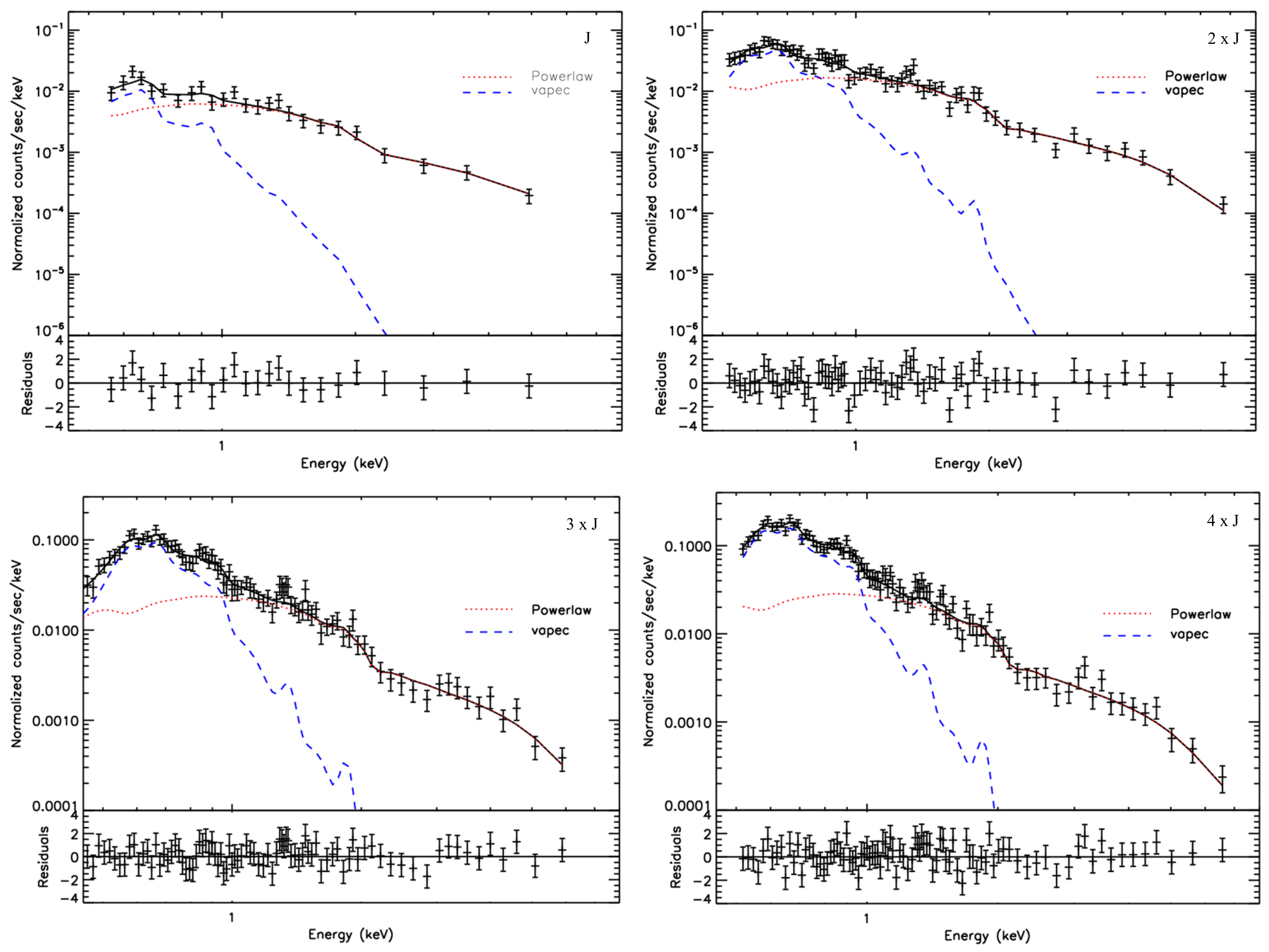}
   \caption{The spectrum from regions surrounding the probable point source location.  The best fit model is shown as a solid black line.  The thermal component is shown as a dashed line with the powerlaw component as a dotted line.}
   \label{pwnspec}
\end{figure}

Finally, region J surrounds the position of a probable point source.  The pulsar position is determined by the \textit{wavdetect} tool using a 2 pixel radius wavelet.  The corresponding $3\sigma$ error ellipse has a physical size of $6.2\times5.3$ pixels.  This area is larger than the \textit{Chandra} point spread function and serves as the most probable location of the pulsar powering the PWN.  As opposed to all the other regions, spectral fits to J show a significant contribution from a non-thermal component.  The best fit model is an absorbed single temperature equilibrium component added to a power law as shown overplotted on the data in Figure \ref{pwnspec}. The thermal fit parameters include a temperature of $kT=0.24^{+0.06}_{-0.04}$ keV with enriched abundances in O and Ne, $4\pm1$ and $<7$ relative to solar, respectively.  The power law describing the high energy emission has a spectral index of $\alpha=1.0\pm0.2$.  Region J does not fully encompass the apparent emission at the core of the remnant.  We therefore investigate the PWN with successively larger extraction regions.  Spectral fits were made for regions encompassing 1.5, 2, 3, and 4 times the area of J.  These regions are ellipses centered on the center of J with the same aspect ratio of semi-major to semi-minor axes.  Figure \ref{Jpic} shows their extent (dashed ellipses) relative to the high energy (1.2--8.0 keV) emission indicative of the PWN.  These raw data are unbinned and gaussian smoothed with a 3 pixel kernel.  There are 10 contours evenly spaced between 0--3 counts per bin.  Table \ref{Jregs} gives the best fit parameters for all of the J regions and spectra with best fit models are shown in Figure \ref{pwnspec}. 

\section{Discussion}
\subsection{Soft X-ray Morphology}
Upon examination of the fit parameters it becomes obvious that SNR 0453-68.5 is dominated by ISM emission outside of the central PWN containing region.  In general, the elemental abundances match those expected from the LMC.  The only elements found to vary are O, and Fe.  Significant departures from accepted LMC values are seen in regions A ($>5\sigma$), E ($>5\sigma$), G ($4\sigma$), and H ($>5\sigma$) as a depletion of oxygen while region E is also underabundant in iron ($>3\sigma$).  There is no direct evidence that any of the regions outside of the PWN display significant ejecta emission.  The departure from the LMC value of oxygen may not be as significant as suggested in Figure \ref{contours}.  When deriving abundances for the \citet{RD1992} paper, the authors used data from an earlier survey of \ion{H}{2} regions and SNR in the Magellanic Clouds \citep{RD1990}.  Both classes of objects displayed large variations in oxygen abundance.  In \ion{H}{2} regions the value was found to be $0.34^{+0.22}_{-0.14}$ relative to solar with $1\sigma$ errors.  Furthermore, N4A, the closest sampled \ion{H}{2} region to SNR 0453-68.5, is on the lower end of that range with an oxygen abundance of 0.22.  When surveying SNR the authors found a relatively lower value of $0.26^{+0.20}_{-0.12}$ ($1\sigma$).  This is consistent with the LMC SNR spectral analysis performed by \citet{Hughes98} using \textit{ASCA} data.  A sample of seven remnants in the LMC, including SNR 0453-68.5, display reduced oxygen abundances averaging $0.26\pm0.05$ (the value found for SNR 0453-68.5 is $0.24\pm0.07$, $90\%$ confidence).  These authors find no evidence for ongoing dust destruction and suggest that the overall reduced abundances found in LMC remnants must reflect the actual abudances of the ISM. Therefore, the depleted oxygen abundance we find here may simply be consistent with LMC SNR in general, just on the lower end.  \citet{Hughes98} find that SNR 0453-68.5 has the lowest metallicity of the sampled SNR across the board suggesting that the local environment is itself depleted, which is supported by the \citet{RD1990} findings on N4A.  In fact, SNR 0453-68.5 lies in a region of the LMC that is somewhat devoid in tracers of star formation such as OB associations, CO emission, and \ion{H}{2} regions \citep{Hughes98}.  This lack of star formation may be responsible for the overly depleted abundances seen in SNR 0453-68.5.

Most of the best fit models contain an equilibrium plasma.  The state of CIE suggests that the remnant is either very old or the emission is dominated by shocked material in over-dense regions.  The latter is often seen in SNR that result from a cavity explosion \citep{McCraySnow}.  The massive progenitor star has a strong wind which blows a cavity in the ISM.  The SN initially sends a blast wave through this vacated volume but then eventually catches up to the cavity wall.  The wall consists of a highly anisotropic, clumpy medium with the densest clumps quickly reaching CIE.  In these regions the ejecta quickly mixes with the dominant, shocked ISM.  This scenario matches the observation with bright filaments of emission near the limb, most notably in regions A, G, and H.  These contain dense regions of the cavity wall that have been shock heated by the SN blast wave.  In contrast, the diffuse, yet bright, interior emission of regions such as B, C, D, and F appear closer to the center of the remnant.  The presence of bright filaments in these regions may be due to line of sight confusion where the most intense emission is still coming from dense areas of the cavity wall but in projection.  This is supported by the fact that these regions are best fit with multiple temperature components; a low temperature equilibrium plasma that has cooled in the dense cavity wall, and a higher temperature non-equilibrium plasma from interior, more rarefied gas that has been reheated by a reflected shock from the cavity wall.  This hotter interior gas is evident in regions E and I.  Similar interaction geometries occur for other shell-type SNR such as the Cygnus Loop \citep{McEntaffer2011}.  The limb brightened morphology could also be due to the concentration of swept up matter over time by the supernova blast wave.  This scenario is supported by the observation that hot interior gas as seen in regions E, and most notably I, is also ISM dominated.  If the intial explosion expanded into a cavity then we might expect that interior emission originates from ejecta.  However, we see no signs of ejecta dominated plasma in these regions.

\begin{figure} [htbp]
   \centering
   \includegraphics[width=5.0in,height=6.0in]{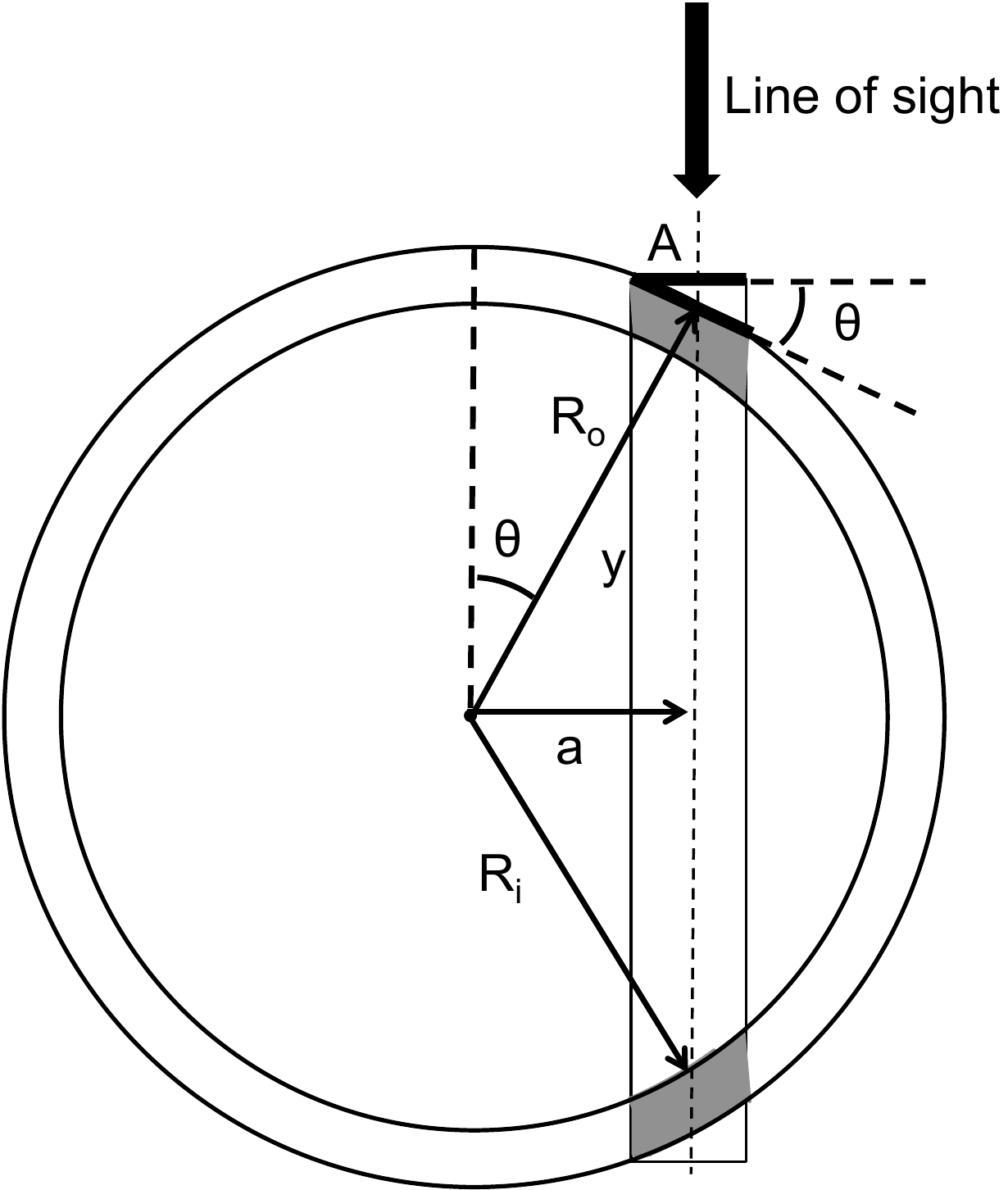}
   \caption{Diagram depicting the geometry used for calculation of the emitting volume (shaded gray) in each region.  The definitions of the labels are as follows: $A$ is the projected area of the region on the sky, $R_o$ is the outer radius of X-ray emission, $R_i$ is the inner radius, $a$ is the projected radial distance to the center of the region, and $\theta$ is the azimuthal offset of the region from the center of the remnant.}
   \label{volume}
\end{figure}

Calculations of shock velocity and density support this conclusion as shown in Table \ref{densityTbl}.  Solving the Rankine-Hugoniot relations in the strong shock case with $\gamma = 5/3$ gives the post-shock temperature as a function of shock velocity, $kT=(3/16) \mu m_{p} v^2$, with $\mu = 0.6$.  We assume that $T_e\sim T_{ion}$ as expected through Coulomb collisions in a plasma with a temperature of $\sim10^6$ K and $n_e\sim1$ cm$^{-3}$.  Such a plasma can equilibrate within a few hundred years or even more rapidly given higher densities (see eqn. 36.37 in \citet{Draine}).  If the electrons and ions are not in temperature equilibrium then the calculated velocities signify lower bounds.  Our resulting shock velocities are typically a few hundred km/s, showing a significant deceleration from the initial blast wave velocity ($>5000$ km/s) and signifying a dense medium.  This density is found using the \textit{norm} parameter for each component, where $norm=[10^{-14}/(4 \pi D^2)] \int n_e n_H dV$. The integral contains the emission measure for the plasma which is dependent on the density and total emitting volume.  A value of 50 kpc is used for $D$ and we assume that $n_e=1.2n_H$.  The volume is calculated for each region assuming that the emission is confined within an annulus.  It is within this annulus that the SN blast wave has interacted with the surrounding medium and it is assumed that the bulk of the X-ray emission, especially for the CIE plasmas, lies within this volume.  A contribution from high temperature, low density gas from the interior of the remnant may exist, but this component is not observed to be significant from the spectral fits.  If this component is indeed a significant contribution to the higher temperature \textit{vnei} components then the densities calculated for these plasmas can be considered as upper limits.  As depicted in Figure \ref{volume}, the volume calculation is then $V = 2Af(\sqrt{R_o^2 - a^2} - \sqrt{R_i^2 - a^2}\ )/cos\theta$, where $A$ is the projected area of the region on the sky, $R_o$ is the outer limit of the X-ray emission (16.8 pc), $R_i$ is an estimated inner radius of the emission annulus (15.8 pc), $a$ is the radial distance to the center of the region, $\theta$ is the azimuthal offset of the region from the center of the remnant, and $f$ is the filling factor of the X-ray emitting gas which ranges from 0 to 1.  Using these calculated volumes, the resulting densities range from $1.9/f-7.6/f$ cm$^{-3}$ in the equilibrium plasmas indicating that dense regions of interaction are responsible for the X-ray emission.  In the regions with multiple temperature components (B, C, D, and F) the equilibrium plasma is always at lower temperature and higher density than the nonequilibrium component.  This is consistent with the above scenario where the SN blast wave interacts with a dense cloud near the limb of the remnant.  The shock decelerates in the density enhancement and equilibrium is reached quickly.  A reverse shock propagates back into the lower density interior reheating this gas to higher temperatures and nonequilibrium conditions.  

We can also estimate the age of the remnant from the equilibrium shock velocities.  Assuming that the remnant can be described by a Sedov solution \citep{Spitzer}, the time since shock heating can be calculated as $t=(2/5)R/v$, where $R$ is the estimated shock distance from the center, 16.8 pc.  As shown in Table \ref{densityTbl} this results in age estimates between 17200-23600 years, significantly older than the 13000 year age used in \citet{Lopez2011} and \citet{Gaensler}.  The spectral fits support an older age for the remnant given that we see no evidence for synchrotron emission near the limb, no evidence for fast shocks, and no evidence for ejecta dominated emission (outside of the J regions), all of which are typically seen in young SNR.  Given the Type II identification of the source there could be leftovers of the molecular cloud that formed the precursor star.  The limb brightened X-ray morphology, ISM dominated emission, and the middle-aged nature of this remnant suggest that the X-ray emission results from the interaction of the SN blast wave with this cloud.  

\begin{figure} [htbp]
   \centering
   \includegraphics[width=6.0in,height=3.0in]{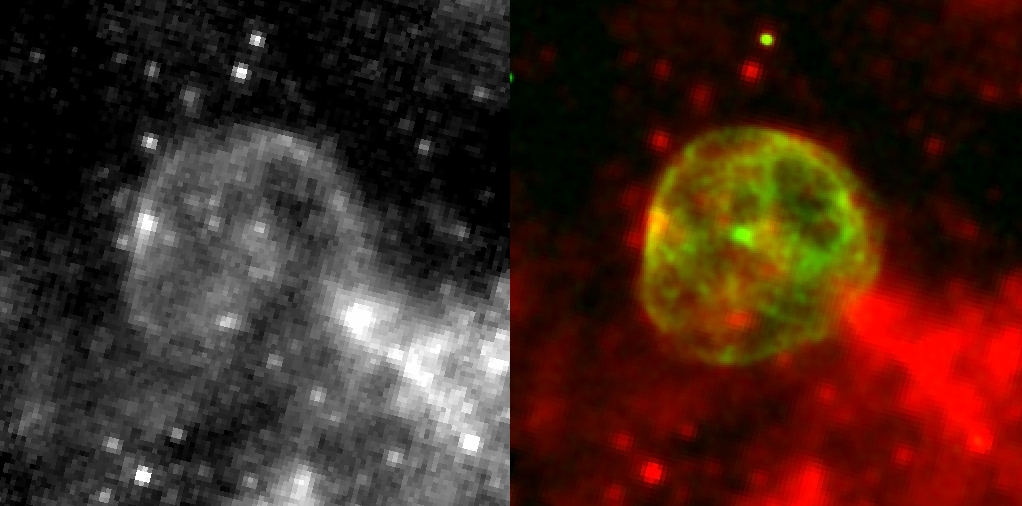}
   \caption{The image on the left is the raw Spitzer data at $24\mu\mathrm{m}$ linearly scaled from 16.35-16.84 counts per bin with no binning or smoothing.  On the right, these data are shown in red and superimposed on the X-ray data in green.}
   \label{Spitzer}
\end{figure}

To further investigate this possibility we reanalyzed $24\mu\mathrm{m}$ \textit{Spitzer} data as shown in Figure \ref{Spitzer}.  \citet{Williams2006} have previously analyzed \textit{Spitzer} data at $24\mu\mathrm{m}$ and $70\mu\mathrm{m}$.  They attribute the emission to warm dust radiating a thermal continuum.  This is consistent with our assumption that the limb brightened X-ray emission is due to interactions of the SN blast wave with dense ISM clouds.  We then use this assumption to compare the morphologies in the X-ray and infrared (IR).  Figure \ref{Spitzer} show the data sets superimposed on one another.  The image on the left is the raw IR data tracing the warm dust in the shock heated ISM.  The image on the right shows the X-ray data in green and the IR data superimposed in red.  Generally, the IR data are well correlated with the X-ray data.  The shell morphology of the remnant is easily noticeable in the IR with enhancements occurring in the same vicinities as those in the X-ray.  The long, smooth shocked region in A appears to be a significant wall of material which shows up in the IR.  There appear to be more discrete IR features in the vicinity as well, but these could be due to line of sight source confusion.  The high density clumps found in the X-ray emission of region B also appear in the IR data.  These bright, dense features in regions A and B are consistent with the equilibrium conditions found from their X-ray spectra.  Filaments seen in the X-ray, such as those in the north corresponding to regions C and F are also present in the IR.  The only region where X-ray emission is bright but IR emission is lacking is the central core of emission surrounding the PWN.  Also, the bright IR cloud in the southwest does not appear to correlate well with X-ray emission.  There is X-ray emission present, but it shows no sign of interaction with a cloud such as a large scale indentation at the limb, a bright knot of emission, or a distinct lack of X-rays due to absorption.  Therefore, it is assumed that this IR feature lies behind the SNR and is not physically related. 

The IR limb brightening noticeable around most of the remnant suggests a nearly spherical distribution of the emitting clouds, thus supporting the idea of swept-up ISM or cavity explosion. The IR data at the limb tend to be outside (at larger radii) the X-ray emission indicating that the shock has preferentially heated the inner parts of these clouds.  A lack of X-ray emission at slightly larger radii could be due extinction in the dense cloud or due to shock deceleration in the deeper recesses of the cloud leading to velocities that are too low to produce X-ray emission.  In addition, there appears to be a general lack of IR emission in the south-southwest region of this remnant.  When compared to the northern hemisphere of the remnant, the south is more dim in the IR.  This relation is mirrored in the X-ray where the northern hemisphere also contains most of the emission (see Figure \ref{raw_xray}).  If we assume that the plerion marks the origin of the SN blast wave then this distribution of IR$/$X-ray emission seems reasonable.  For instance, the eastern limb is somewhat closer to the plerion than the western limb, but this is probably due to the larger, denser cloud in the east slowing the shock more.  Also, the plerion appears to be located north of the center of emission.  However, the low density in the south and southwest has allowed the blast wave to propagate further from the explosion origin thus increasing the distance from the limb to the center in comparison to the north.  Therefore, the off-center position of the plerion may not be due to an asymmetric explosion imparting velocity to the neutron star.  Instead, this analysis of the IR morphology would argue that the pulsar has very little proper motion relative to the remnant, especially when considering the $>$17000 year age.  The pulsar may be located very close to the original explosion origin unless its direction of motion is along the line of sight.

The fact that SNR 0453-68.5 is an older, ISM dominated remnant is an interesting result because the only other previous study on the shell X-ray morphology of this remnant, \citet{Lopez2009,Lopez2011}, could not discern between explosion or environment as the cause of the asymmetry.  The Lopez et al. studies use a ``power ratio method'' which measures X-ray surface brightness asymmetries using multipole moments seen in the soft X-ray (0.5-2.1 keV) and in \ion{Si}{13} (data between 1.75--2.0 keV).  A comparison of the quadrupole versus the octupole ratios of the emission allow the authors to categorize 24 SNR in the Milky Way and LMC into two distinct groups which are consistent with their explosion type - Type Ia versus core collapse (CC).  SNR 0453-68.5 is quite close to the dividing line in the broadband X-ray analysis.  It is more distinctly classified as a CC remnant when considering \ion{Si}{13} emission.  Therefore, their result is that the distribution of X-ray surface brightness, which as an initial condition is presumed to be dominated by ejecta emission, leads to a CC classification for SNR 0453-68.5.  We agree with the classification of this object and build on their result by showing that in the case of this particular Type II SNR, the morphology is not determined by the asymmetric distribution of ejecta arising from an asymmetric CC explosion, but rather inhomogeneity in the surrounding ISM. When we extract data from the 1.75--2.0 keV energy range, which is where the dominant \ion{Si}{13} triplet lies, we see a surface brightness distribution that matches the low energy X-rays.  These data trace the interaction at the limb of the remnant which is dominated by the ISM, not clumps of ejecta.  Furthermore, given the near constant position of the pulsar, the distribution of ISM seen in the IR, and the nearly symmetric blast wave propagation (after consideration of the cloud distribution suggested from the IR) it appears that the explosion that created SNR 0453-68.5 was quite likely very symmetric.   Any evidence from the ejecta has been lost through mixing as seen from the ISM dominated spectral fits.  There appears to be no evidence from this remnant that the explosion mechanism that led to this CC SNR was inherently asymmetric.

\subsection{PWN/Reverse Shock Interaction}
The pulsar and its associated emission are detected in region J.  The spectral index of the high energy emission, $\alpha=1.0\pm0.2$, is consistent with the range from other remnants containing a PWN in the LMC: $\alpha=0.4-1.4$ for a series of annular regions encompassing the PWN of SNR 0540-69.3 \citep{Petre}; $\alpha=1.2\pm0.3$ in N206 (SNR 0532-71.0) \citep{Williams}; $\alpha=0.57^{+0.05}_{-0.06}$ for DEM L241 \citep{Bamba}; and $\alpha\sim1.2-1.6$ over several regions from N157B \citep{Chen}.

Furthermore, it is consistent with a previous fit quoted in \citet{Gaensler}, $\alpha=0.9\pm0.4$.  However, \citet{Gaensler} do not find the presence of a point source to be significant.  To test the presence of a point source they model the central emission using a combination of an elliptical Gaussian, a point source, and background.  They then state that the fit is equal or better if the point source is resolvable, thus leading to a conclusion that there is no strong detection of a point source.  However, \citet{Lopez2011} also search for a pulsar in SNR 0453-68.5 and find a ``bright pulsar'' using \textit{wavdetect}, consistent with the findings here.  Even without a point source detection the presence of a pulsar is likely given the detected synchrotron nebula which is powered by the embedded pulsar.

SNR 0453-68.5 exhibits a morphology that is consistent with a composite remnant.  It has a thermal shell of emission from interaction with the ambient ISM and a central nebula of synchrotron emission surrounding a pulsar.  Theoretical evolution of a PWN suggests that the expanding nebula will inevitably interact with a reverse shock (RS) from the SNR \citep{GaenslerSlane}.  The RS is formed during the Sedov phase of SNR evolution when the forward shock has swept up significant ISM mass in comparison to the ejecta mass thus forming a contact discontinuity.  The RS will eventually propagate inward and reach the center of the remnant within several thousand years.  However, it will encounter the PWN before this time resulting in compression of the nebula.  The PWN will react with an increase in pressure and subsequent expansion.  This battle continues for several thousand years resulting in complex morphology and mixing at the unstable interface.  This mixing will occur between the synchrotron emitting plasma of the PWN and the hot, shocked, thermally emitting ejecta near the interface.  Any asymmetry in these variables such as ISM distribution, RS velocity, or pulsar velocity adds significant additional complexity.  The obvious presence of the PWN in SNR 0453-68.5 and its age of $>17000$ years suggest that this interaction has or is currently taking place.  The best fit spectrum from region J supports this conclusion (Figure \ref{Jregs}).  The spectrum shows the presence of obvious emission lines at soft energies, most notably \ion{O}{8} and \ion{Ne}{9}, which leads to a significant thermal contribution to the fit.  Furthermore, the abundances determined from this fit are significantly enhanced relative to the local ISM ($\mathrm{O}=4\pm1$ and $\mathrm{Ne}<7$ relative to solar) suggesting that the plasma is ejecta dominated.  Therefore, the spectral properties of region J combined with the expected evolutionary stage of the PWN, given a 17000 year age, suggest that we are observing this remnant during the interaction of the PWN with the RS of the SNR.

Inspection of Figure \ref{Jpic} gives hints at complexities within the PWN.  It is somewhat elongated along the northeast to southwest direction.  There also appears to be a clump of emission located to the south-southwest of the point source.  Although these features are intriguing, a longer \textit{Chandra} observation would be necessary for investigating the structure in more detail.  Given the short observation, we are not able to perform detailed extractions to exhaustively probe this RS/PWN interaction.  However, using regions that are 1.5$\times$, 2$\times$, 3$\times$, and 4$\times$ the area of region J we are still able to investigate the interaction and make comparisons to other PWN.

There are several trends that are noticeble from the spectral fits to the regions surrounding the point source (see Table \ref{Jregs}).  First, the abundances of O and Ne remain enriched relative to the expected abundances in the LMC, which is consistent with the region being ejecta dominated.  Second, as the size of the region increases the relative contribution from the power law component decreases in comparison to the thermal component.  This is interpreted as an increase in the filling factor of hot, shocked, thermal ejecta as the regions encompass more of this plasma relative to the confined PWN emission closer to the center.  It appears that the relative contributions become equal just above a region size of 1.5$\times$J.  Next, the photon index remains constant at $\Gamma\sim2$.  If we assume that the density of the PWN is constant, then the increasing $norm$ is due to the increased volume of nebula encompassed with each region.  Therefore, subsequent regions sample nonthermal plasma at increasing radii from the pulsar, although the modest increase in the synchrotron emitting volume covers a radial increase of less than a factor of 2 in the PWN plasma.  The constant photon index suggests that the distribution of nonthermal particles remains the same throughout the regions probed, i.e. the electron distribution is constant over these radii given $\Gamma=\alpha+1$ and $p=2\alpha+1$ where p is the power law index for the electron distribution.  At the LMC distance of 50 kpc the semi-major axes of these ellipses probe scales from 0.4-1.2 pc.  In contrast, other PWN containing remnants show a significant change in the photon index over these scales.  The Crab nebula varies from $\Gamma=1.8-2.4$ over a 1.2 pc scale \citep{Willingale} while SNR 0540-69.3 has a variation of $\Gamma=1.4-2.4$ over 1.2 pc as well \citep{Petre}.  Both remnants show a spectral hardening at smaller radii which is attributed to a radial dependence on the magnetic field.  It is important to note that at this scale these remnants are dominated by the PWN emission versus thermal emission.  In SNR 0453-68.5 we see a substantial contribution ($>22\%$) from thermal plasma even at the smallest scales.  Therefore, even though we probe to a similar spatial scale we see a much smaller physical volume dominated by the PWN in comparison to other remnants.  We believe that this can be attributed to the age of SNR 0453-68.5.  The interaction between the PWN and the RS has matured during the $>$17000 year timescale.  Over this time efficient mixing has occured between the hot ejecta and the synchrotron plasma.  If there is a PWN dominated volume then it is contained within 0.4 pc.  Outside of this the oscillations of the PWN/RS interaction have mixed the thermal and non-thermal plasmas.  The constant temperature displayed by these regions may be indicative of this as well.  Any variation in $\Gamma$ or spectral hardening would exist on too fine of a spatial scale to be determined from this observation.

Another LMC PWN, contained within DEM L241, displays a similarly limited spatial extent as SNR 0453-68.5.  \citet{Bamba} observed DEM L241 with \textit{XMM-Newton} and resolved X-ray structures that consist of a hard central source surrounded by diffuse emission along with a tail of extended diffuse emission.  The hard central source is determined to be a PWN well fit by a power law with $\Gamma=1.57$.  The extent of the central source is consistent with the \textit{XMM} PSF of 5'' which translates to 1.2 pc at 50 kpc.  Therefore, determining variability of the spectral index is limited by the spatial resolution.  However, the authors determine that this remnant is very old, $>1$ Myr, which would suggest that any RS/PWN interaction would have taken place long ago and input power from the pulsar has significantly decreased.  Furthermore, the situation for DEM L241 is quite different given its environment.  Despite its old age, the diffuse emission from this remnant is ejecta dominated and centrally filled; it lacks any shell of emission.  This may result from the fact that DEM L241 is contained within an OB association.  The massive star winds have created a large, low density bubble in which the SNR is expanding.  The lack of ISM dominated shell emission suggests that a reverse shock is weak or non-existent, thus complicating the evolution story for the DEM L241 PWN and making connections with SNR 0453-68.5 difficult.  Similar to DEM L241, N157B, another LMC PWN containing SNR, is located within a massive star association, but exhibits a resolved RS/PWN interaction \citep{Chen}.  In the case of N157B the PWN exhibits a cometary shape with a pulsar power law index of $\Gamma=1.73$, a surrounding nonthermal coma with a power law index of $\Gamma=2.18$ and a nonthermal tail with $\Gamma=2.62$.  \citet{Chen} put forth the possibility that the shape of the nebula is not determined by pulsar proper motion but rather by a reverse shock propagating from a dense cloud close to the explosion.  The PWN and SNR have expanded freely in the low density medium close to the massive stars, but are now encountering the edges of the bubble in this very complicated star formation region.  Optical and IR data reveal the presence of a probable cloud responsible for this reverse shock.  Therefore, there are many unique morphologies between the LMC PWNe, all of which are highly influenced by the local environment.  SNR 0453-68.5 is no exception to this and is a member of a class characterized by middle-aged, composite remnants with evolved RS/PWN interactions. 

The morphology of the X-ray and infrared radiation as well as the central location of the detected point source indicate that there is little to no proper motion of the SNR 0453-68.5 pulsar transverse to the line of sight.  Further evidence is the lack of a relic PWN. If the pulsar experienced a typical kick from the SN of a few hundred km/s then we would expect it to be $\sim25\%$ of the way to the limb.  It would generate a new PWN in its current position which would be observable in X-rays while leaving behind the original nebula which would still glow in the radio, hence the relic PWN.  Therefore, any proper motion should show a displacement between the X-ray and radio emission.  However, \citet{Gaensler} show that the radio and X-ray centroids coincide.  This lack of offset, the $>$17000 year age, and the mature PWN/RS interaction all argue that the pulsar could still be centrally located with very little, if any, velocity imparted from the SN.  An LMC PWN that exhibits a pulsar with significant proper motion is N206.  \citet{Williams} performed a combined X-ray, optical, and radio analysis of this remnant.  They classify N206 as a mixed-composite SNR because it exhibits the composite nature with a PWN and shell while also showing a mixed morphology with central X-rays not associated with the PWN.  They find a compact hard X-ray source near the limb of emission at the tip of an elongated radio feature.  They interpret this as a pulsar with significant proper motion moving with respect to the SNR.  The radio feature is the trail of synchrotron emission in the pulsar's wake and a bow shock structure in the X-ray is also evident.  This morphology is in stark contrast to that shown by SNR 0453-68.5, which shows none of these features.

\section{Summary}
We have presented the first in-depth analysis of the X-ray spatial and spectral properties of SNR 0453-68.5 in the LMC.  This is a composite remnant containing a limb brightened shell and a central PWN.  We find that this remnant is middle-aged with ISM dominated X-ray emission.  We also find that any X-ray emission external to the PWN location can be characterized by a one or two temperature component plasma.  The single temperature fits are either low temperature equilibrium plasmas seen as dense, shocked filaments at the edge of the remnant, or high temperature nonequilibrium plasmas in the lower density interior.  Two temperature fits consist of a mix of these two components due to line of sight confusion for the interior regions.  The interpretation is that the morphology is due to ISM swept up by the SN blast wave or due to a cavity explosion.  In either case, interactions within the dense cavity wall lead rapidly to CIE conditions.  A reverse shock from the cavity wall may propagate inward reheating the more rarefied plasma interior to the wall thus contributing the NEI plasma.  This scenario is supported by the following findings: 1) the IR emission is correlated with the X-ray emission and shows a swept-up shell of dense material; 2) this correlation is due to density enhancements in the cavity wall; 3) the spectral fits have elemental abundances consistent with LMC ISM and show no sign of ejecta; 4) the age of the remnant is most likely $>$17000 years; 5) the presence of centralized synchrotron emission, most likely surrounding a compact source, verifies the Type II origin and thus the possible presence of a leftover molecular cloud and/or a strong precursor stellar wind.  

Analysis of the central source shows the clear detection of a power law component due to the PWN emission.  The photon index, $\Gamma\sim2.0$, is consistent with other LMC PWN.  The characteristics of this plasma and an associated thermal plasma remain constant over a 0.4-1.2 pc scale indicating significant mixing of the plasma.  These spectral fits suggest that the evolution of this PWN is currently deep within the PWN/RS interaction phase.  The timescale for this interaction is consistent with the calculated age of the remnant.  The central source exhibits no sign of having a significant velocity transverse to the line of sight.  There is no evidence of a relic PWN, and the apparent offset of the central source from the geometric center of emission is caused by density variations in the molecular cloud wall.  In total, our results show that SNR 0453-68.5 is a unique example of an evolved composite remnant showing clear signs of a PWN/RS interaction.

\section{Acknowledgments}\label{ack}
The authors would like to acknowledge internal funding initiatives at the University of Iowa for support of this work.  Thomas Brantseg is supported by NASA grant NNX10AN16H.  The data used here were obtained from the Chandra Data Archive.

\clearpage


\begin{deluxetable} {c c c c c c c c c c}
 \tablewidth{7 in}
 \tabletypesize{\scriptsize}
 \rotate
 \tablecaption{Best fit parameters for spectral extraction regions}
 \tablehead{
  \colhead{Region} &
  \colhead{$\chi ^2$/$\nu$} &
  \colhead{$kT_{vnei}$} &
  \colhead{$kT_{vapec}$} &
  \colhead{O} &
  \colhead{Fe} &
  \colhead{$\tau$} &
  \colhead{$norm_{vnei}$} &
  \colhead{$norm_{vapec}$} &
  \colhead{$N_H$} \\
  & & (keV) & (keV) & & & ($10^{11}$ s cm$^{-3}$) & $10^{-3}A$\tablenotemark{a} & $10^{-3}A$\tablenotemark{a} & ($10^{22}$ cm$^{-2}$)
 }
 \startdata
 A & 25/43 & \nodata & $0.181^{+0.006}_{-0.004}$ & $0.17^{+0.03}_{-0.02}$ & \nodata & \nodata & \nodata & $1.5^{+0.5}_{-0.4}$ &  $0.14 \pm 0.03$\\
 B & 62/51 & $0.28^{+0.02}_{-0.03}$ & $0.10\pm0.02$	& \nodata & \nodata & $>6$ & $0.7^{+0.4}_{-0.2}$ & $4^{+13}_{-2}$ & $0.16^{+0.07}_{-0.02}$\\
 C & 81/55 & $0.26^{+0.01}_{-0.03}$ & $0.11^{+0.01}_{-0.05}$ & \nodata & \nodata &	$>3$ & $1.0^{+1}_{-0.2}$ & $4^{+34}_{-2}$ & $0.16^{+0.07}_{-0.02}$ \\
 D & 79/55 & $0.9\pm0.1$ &	$0.18^{+0.02}_{-0.01}$ & $0.18^{+0.11}_{-0.04}$ & $0.8^{+0.4}_{-0.2}$ & $>10$ & $0.09\pm0.02$ & $2^{+2}_{-1}$ & $0.16^{+0.07}_{-0.08}$\\
 E & 76/59 & $0.35\pm0.02$ & \nodata & $0.16^{+0.02}_{-0.01}$ & $0.36 \pm 0.04$ & $1.0^{+0.2}_{-0.5}$ & $0.6 \pm 0.1$ & \nodata & $0.08^{+0.02}_{-0.03}$\\
 F & 61/49 & $0.24^{+0.09}_{-0.01}$ & $0.10^{+0.04}_{-0.02}$ & \nodata	& \nodata & $23\pm4$ & $9^{+3}_{-6}$ & $4^{+5}_{-2}$ &	$0.17^{+0.03}_{-0.02}$\\
 G & 32/34 & \nodata & $0.18 \pm 0.01$ & $0.17^{+0.04}_{-0.03}$ & \nodata & \nodata & \nodata & $0.97^{+0.06}_{-0.03}$ & $0.14^{+0.04}_{-0.03}$\\
 H & 45/42 & \nodata & $0.18 \pm 0.01$ & $0.15^{+0.03}_{-0.02}$ & $0.3 \pm 0.1$ & \nodata & \nodata & $1.7^{+0.7}_{-0.5}$ & $0.16^{+0.04}_{-0.03}$\\
 I & 61/40 & $0.42^{+0.07}_{-0.05}$ & \nodata & $0.25^{+0.05}_{-0.04}$ & \nodata & $0.5\pm0.3$ & $0.11^{+0.07}_{-0.05}$ & \nodata	& $0.10 \pm 0.05$ \enddata
 \tablecomments{Abundances are given relative to solar, with solar values = 1.  Only variable abundances are shown with the rest held at LMC values.}
 \tablenotetext{a}{Normalization parameter, where $A=[10^{-14}/(4\pi D^2)]  \int n_e n_H dV$. D is the distance to the LMC and the integral is the volume emission measure.}
 \label{fitparams}
\end{deluxetable}

\clearpage


\begin{deluxetable} {r c c c c c c}
 \tablecaption{Best fit parameters for J and surrounding regions. \label{bigtable}}
 \tablehead{
  \colhead{Region} &
  \colhead{$kT$} &
  \colhead{O} &
  \colhead{Ne} &
  \colhead{$\Gamma$} &
  \colhead{$norm_{vapec}$} &
  \colhead{$norm_{plaw}$}\\
  & (keV) & & & & $10^{-5} A$ & $10^{-5} A$
 }
 \startdata
 J & $0.24^{+0.06}_{-0.04}$ & $4 \pm 1$ & $<7$ & $2.0 \pm 0.2$ & $0.5 \pm 0.1$ & $1.7 \pm 0.2$\\
 1.5$\times$J & $0.23^{+0.3}_{-0.2}$ & $1.8 \pm 0.3$ & $<2.6$ & $2.1 \pm 0.1$ & $2.3 \pm 0.3$ & $3.3 \pm 0.2$\\
 2$\times$J & $0.22 \pm 0.01$ & $1.0 \pm 0.1$ & $0.7 \pm 0.4$ & $2.0 \pm 0.1$ & $8.1 \pm 0.8$ & $4.4 \pm 0.3$\\
 3$\times$J & $0.21 \pm 0.01$ & $0.52 \pm 0.04$ & $0.7 \pm 0.2$ & $2.03 \pm 0.08$ & $32 \pm 2$ & $6.3 \pm 0.3$\\
 4$\times$J & $0.204 \pm 0.004$ & $0.48 \pm 0.03$ & $0.71 \pm 0.1$ & $2.07 \pm 0.08$ & $59 \pm 3$ & $7.4 \pm 0.4$
 \enddata
 \label{Jregs}
\end{deluxetable}

\clearpage


\begin{deluxetable} {c c c c c c c}
 \tablecaption{Calculated region parameters}
 \tablehead{
  \colhead{Region} &
  \colhead{v$_{vnei}$} &
  \colhead{v$_{vapec}$} &
  \colhead{Volume $\times f$} &
  \colhead{$n_{e_{vnei}} / f$} &
  \colhead{$n_{e_{vapec}} /f $} &
  \colhead{Age}\\
  & (km/s) & (km/s) & (pc$^3$) & (cm$^{-3}$) & (cm$^{-3}$) & (yr)
}
 \startdata
 A & \nodata & 383 & 73 & \nodata & 5.0 & 17200\\
 B & 480 & 296 & 130 & 2.3 & 4.5 & 22200\\
 C & 460 & 293 & 130 & 3.1 & 6.4 & 22500\\
 D & 846 & 380 & 94 & 1.1 &	4.9 & 17300\\
 E & 533 & \nodata & 105 & 2.6 & \nodata & \nodata\\
 F & 442 & 278 & 78 & 3.8 & 7.6 & 23600\\
 G & \nodata & 382 & 318 & \nodata & 1.9 & 17200\\
 H & \nodata & 382 & 79 & \nodata & 5.1 & 17200\\
 I & 584 & \nodata & 237 & 0.75 & \nodata & \nodata\\
 \enddata
 \label{densityTbl}
\end{deluxetable}

\clearpage


\clearpage

\clearpage

\clearpage

\clearpage

\clearpage

\clearpage

\clearpage

\clearpage

\clearpage

%


\begin{thebibliography}{}
\bibitem[Anders \& Grevesse(1989)]{angr} Anders, E., \& Grevesse, N. 1989, \gca, 53, 197
\bibitem[Baluncinska-Church \& McCammon(1992)]{bcmc} Balucinska-Church, M., \& McCammon, D. 1992, \apj, 400, 699
\bibitem[Bamba et al.(2006)]{Bamba} Bamba, A., Ueno, M., Nakajima, H., Mori, K., \& Koyama, K. 2006, \aa, 450, 585
\bibitem[Borkowski et al.(1994)]{bork94} Borkowski, K. J., Sarazin, C. L., \& Blondin, J. M. 1994, \apj, 429, 710
\bibitem[Borkowski et al.(2001)]{bork01} Borkowski, K. J., Lyerly, W. J., \& Reynolds, S. P. 2001, \apj, 548, 820
\bibitem[Chen et al.(2006)]{Chen} Chen, Y., Wang, Q. D., Gotthelf, E. V., Jiang, B., Chu, Y., Gruendl, R. 2006, \apj, 651, 237
\bibitem[Dennerl et al.(2001)]{Dennerl} Dennerl, K. et al. 2001, \aa, 365, L202
\bibitem[Draine(2011)]{Draine} Draine, B.~T.\ 2011, Physics of the Interstellar and Intergalactic Medium by Bruce T.~Draine.~Princeton University Press, 2011.~ISBN: 978-0-691-12214-4 
\bibitem[Foster et al.(2011)]{aped2} Foster, A. R., Smith, R. K., \& Brickhouse, N. S. 2011, in press
\bibitem[Freeman et al.(2001)]{Sherpa} Freeman, P., Doe, S., \& Siemiginowska, A. 2001, \procspie, 4477, 76
\bibitem[Freeman et al.(2002)]{freeman} Freeman, P. E., Kashyap, V., Rosner, R., Lamb, D. Q. 2002, \apjs, 138, 185
\bibitem[Fruscione et al.(2006)]{CIAO2006} Fruscione, A., et al. 2006, \procspie, 6270, 62701V
\bibitem[Gaensler et al.(2003)]{Gaensler} Gaensler, B. M., Hendrick, S. P., Reynolds, S. P., \& Borkowski, K. J. 2003, \apj, 594, L111
\bibitem[Gaensler \& Slane(2006)]{GaenslerSlane} Gaensler, B. M., \& Slane, P. O. 2006, \araa, 44, 17
\bibitem[Hamilton et al.(1983)]{Hamilton} Hamilton, A. J. S., Chevalier, R. A., \& Sarazin, C. L. 1983, \apjs, 51, 115
\bibitem[Hughes et al.(1998)]{Hughes98} Hughes, J. P., Hayashi, I., \& Koyama, K. 1998, \apj, 505, 732
\bibitem[Kalberla et al.(2005)]{LAB2005} Kalberla, P. M. W., Burton, W. B., Hartmann, D., Arnal, E. M., Bajaja, E., Morras, R. \& Poeppel, W. G. L. 2005, \aa, 440, 775
\bibitem[Kargaltsev \& Pavlov(2008)]{KP2008} Kargaltsev, O., \& Pavlov, G. G. 2008, AIPC, 983, 171
\bibitem[Liedahl et al.(1995)]{liedahl95} Liedahl, D. A., Osterheld, A. L., Goldstein, W. H. 1995, \apjl, 438, L115
\bibitem[Lopez et al. (2009)]{Lopez2009} Lopez, L. A., Ramirez-Ruiz, E., Badenes, C., Huppenkothen, D., Jeltema, T. E., \& Pooley, D. A. 2009, \apj, 706, L106
\bibitem[Lopez et al.(2011)]{Lopez2011} Lopez, L. A., Ramirez-Ruiz, C., Huppenkothen, E., Badenes, \& Pooley, D. A. 2011, \apj, 732, 114
\bibitem[Mazzotta et al.(1998)]{mazz} Mazzotta, P., Mazzitelli, G., Colafrancesco, S., \& Vittorio, N. 1998, \aaps, 133, 403
\bibitem[McCray \& Snow(1979)]{McCraySnow} McCray, R., \& Snow, Jr., T. P. 1979, \araa, 17, 213
\bibitem[McEntaffer \& Brantseg(2011)]{McEntaffer2011} McEntaffer, R. L., \& Brantseg, T. 2011, \apj, 730, 99
\bibitem[Park et al.(2010)]{Park} Park, S., Hughes, J. P., Slane, P. O., Mori, K., Burrows, D. N. 2010, \apj, 710, 948
\bibitem[Petre et al.(2007)]{Petre} Petre, R., Hwang, U., Holt, S. S., Safi-Harb, S., \& Williams, R. M. 2007, \apj, 662, 988
\bibitem[Reynolds(1998)]{Reynolds1998} Reynolds, S. P. 1998, \apj, 493, 375
\bibitem[Reynolds \& Keohane(1999)]{RK1999} Reynolds, S. P. \& Keohane, J. W. 1999, \apj, 525, 368
\bibitem[Rieke et al.(2004)]{RiekeYoung2004} Rieke, G. H., et al. 2004, \apjs, 154, 25
\bibitem[Russell \& Dopita(1992)]{RD1992} Russell, S.C., \& Dopita, M. A. 1992, \apj, 384, 508
\bibitem[Russell \& Dopita(1990)]{RD1990} Russell, S.C., \& Dopita, M. A. 1990, \apjs, 74, 93
\bibitem[Sedov(1993)]{sedov} Sedov, L. 1993, Similarity and Dimensional Methods in Mechanics (10th ed.; Boca Raton, Fla. : CRC Press)
\bibitem[Smith et al.(2001)]{aped1} Smith, R. K., Brickhouse, N. S., Liedahl, D. A., Raymond, J. C. 2001, \apj, 556, L91
\bibitem[Smith \& Hughes(2010)]{Smith2010} Smith, R. K., \& Hughes, J. P. 2010, \apj, 718, 583
\bibitem[Spitzer(1998)]{spitzer} Spitzer, L. 1998, Physical Processes in the Interstellar Medium (Wiley Classics Library Edition; New York, NY; Wiley Interscience)
\bibitem[Wang \& Gotthelf(1998)]{WG98} Wang, D. Q., \& Gotthelf, E. V. 1998, \apj, 509, L109
\bibitem[Williams et al.(2005)]{Williams} Williams, R. M., Chu, Y.-H., Dickel, J. R., Gruendl, R. A., Seward, F. D., Guerrero, M. A., \& Hobbs, G. 2005, \apj, 628, 704
\bibitem[Williams et al.(2006)]{Williams2006} Williams, B. J., et al. 2006, \apj, 652, L33
\bibitem[Willingale et al.(2001)]{Willingale} Willingale, R., Aschenbach, B., Griffiths, R. G., Sembay, S., Warwick, R. S., Becker, W., Abbey, A. F., \& Bonnet-Bidaud J.-M. 2001, \aa, 365, L212

\end{thebibliography}
\end{document}